\documentclass[aps,floats]{revtex4}
\usepackage{amsmath,amssymb}
\usepackage{graphicx,epsfig}

\begin{document}
\bibliographystyle {plain}

\def\oppropto{\mathop{\propto}} 
\def\opmin{\mathop{\min}} 
\def\opmax{\mathop{\max}} 
\def\opsimeq{\mathop{\simeq}}
\def\opoverderline{\mathop{\overline}}
\def\operarrow{\mathop{\longrightarrow}}
\def\opsim{\mathop{\sim}}

\def\fig#1#2{\includegraphics[height=#1]{#2}}
\def\figx#1#2{\includegraphics[width=#1]{#2}}


\title{ Statistics of the two-point transmission 
at Anderson localization transitions  } 


 \author{ C\'ecile Monthus and Thomas Garel }
\affiliation{Institut de Physique Th\'{e}orique, CNRS and CEA Saclay
 91191 Gif-sur-Yvette cedex, France}

\begin{abstract}
At Anderson critical points, the statistics of the two-point transmission $T_L$ for disordered samples of linear size $L$ is expected to be multifractal with the following properties [Janssen {\it et al} PRB 59, 15836 (1999)] : (i) the probability to have $T_L \sim 1/L^{\kappa}$ behaves as $L^{\Phi(\kappa)}$, where the multifractal spectrum $\Phi(\kappa)$ terminates at $\kappa=0$ as a consequence of the physical bound $T_L \leq 1$; (ii) the exponents $X(q)$ that govern the moments $\overline{ T_L^q} \sim 1/L^{X(q)}$ become frozen above some threshold: $X(q \geq q_{sat}) = - \Phi(\kappa=0)$, i.e. all moments of order $q \geq q_{sat}$ are governed by the measure of the rare samples having a finite transmission ($\kappa=0$). In the present paper, we test numerically these predictions for the ensemble of $L \times L$ power-law random banded matrices, where the random hopping $H_{i,j}$ decays as a power-law $(b/\vert i-j \vert)^a$. This model is known to present an Anderson transition at $a=1$ between localized ($a>1$) and extended ($a<1$) states, with critical properties that depend continuously on the parameter $b$. Our numerical results for the multifractal spectra $\Phi_b(\kappa)$ for various $b$ are in agreement with the relation $\Phi(\kappa \geq 0) = 2 \left[ f( \alpha= d+ \frac{\kappa}{2}   ) -d \right]$ in terms of the singularity spectrum $f(\alpha)$ of individual critical eigenfunctions, in particular the typical exponents are related via the relation $\kappa_{typ}(b)= 2 (\alpha_{typ}(b)-d)$. We also discuss the statistics of the two-point transmission in the delocalized phase and in the localized phase.

\end{abstract}

\maketitle

\section{ Introduction  }

Since its discovery fifty years ago \cite{anderson},
Anderson localization has remained a very active field
of research (see the reviews \cite{thouless,souillard,bookpastur,
Kramer,janssenrevue,markos,mirlinrevue}).
One of the most important property of Anderson localization transitions
is that critical eigenfunctions are described by
 a multifractal spectrum $f(\alpha)$ defined as follows
(for more details see for instance the reviews 
\cite{janssenrevue,mirlinrevue}):
in a sample of size $L^d$, the number ${\cal N}_L(\alpha)$
of points $\vec r$ where the weight $\vert \psi(\vec r)\vert^2$
scales as $L^{-\alpha}$ behaves as 
\begin{eqnarray}
{\cal N}_L(\alpha) \oppropto_{L \to \infty} L^{f(\alpha)}
\label{nlalpha}
\end{eqnarray}
The inverse participation ratios (I.P.R.s) can be then rewritten
as an integral over $\alpha$
\begin{equation}
Y_q(L)  \equiv \int_{L^d} d^d { \vec r}  \vert \psi (\vec r) \vert^{2q}
\simeq \int d\alpha \ L^{f(\alpha)} 
\ L^{- q \alpha} \opsimeq_{L \to \infty} 
L^{ - \tau(q) }
\label{ipr}
\end{equation}
The exponent $\tau(q)$ can be obtained via a saddle-point
calculation in $\alpha$, and one obtains the Legendre
transform formula \cite{janssenrevue,mirlinrevue}
\begin{eqnarray}
 q && =f'(\alpha) \nonumber \\
 \tau(q) && =  q \alpha  - f(\alpha)
\label{legendre}
\end{eqnarray}
These scaling behaviors, which concern  
individual eigenstates $\psi$, can be translated
for the local density of states
\begin{equation}
\rho_L(E,\vec r) = \sum_{n} \delta(E-E_n) \vert \psi_{E_n}(\vec r)\vert^2
\label{defrho}
\end{equation}
as follows : for large $L$, when the $L^d$ energy levels become dense,
the sum of Eq. \ref{defrho} scales as
\begin{equation}
\rho_L(E, \vec r) \propto L^d \vert \psi_E(\vec r)\vert^2
\label{equiv}
\end{equation}
and its moments involve the exponents $\tau(q)$ introduced in Eq. \ref{ipr}
\begin{equation}
\overline{ [\rho_L(E,\vec r)]^q } \oppropto_{L \to \infty}
\frac{1}{L^{\Delta(q)}} \ \ {\rm with \ \ } \Delta(q) =  \tau(q)-d (q-1) 
\label{rhomoments}
\end{equation}
These notions concern one-point functions, and it is natural 
to consider also the statistics of two-point functions.
In particular, a very interesting observable 
to characterize Anderson transitions
is the two-point transmission $T_L$ when the 
disordered sample of size $L^d$ is attached to one incoming wire
and one outcoming wire \cite{janssen99,evers08} :
(i) it remains finite in the thermodynamic limit 
only in the delocalized phase,
so that it represents an appropriate 
order parameter for the conducting/non-conducting transition;
(ii) exactly at criticality, it displays multifractal properties 
in direct correspondence with the multifractality of critical eigenstates,
i.e. it displays strong fluctuations that are not captured by more global
definitions of conductance.
More precisely, as first discussed in \cite{janssen99}
for the special case of the two dimensional quantum Hall transition,
the critical probability distribution of the two-point transmission $T_L$ 
takes the form
\begin{equation}
{\rm Prob}\left( T_L \sim L^{-\kappa}  \right) dT
\oppropto_{L \to \infty} L^{\Phi(\kappa) } d\kappa
\label{phikappa}
\end{equation}
and its moments involve non-trivial exponents $X(q)$
\begin{equation}
\overline{T_L^q} \sim \int d\kappa L^{\Phi(\kappa) -q \kappa }
\oppropto_{L \to \infty} L^{-X(q)}
\label{defXq}
\end{equation}
As stressed in \cite{janssen99}, the physical bound $T_L \leq 1$
on the transmission implies that the multifractal spectrum
exists only for $\kappa \geq 0$, and this termination at $\kappa=0$
 leads to a complete freezing of the moments exponents
 \begin{eqnarray}
X(q)  =X(q_{sat}) \ \ \ \ {\rm for } \ \ q \geq q_{sat}
\label{freezing}
\end{eqnarray}
at the value $q_{sat}$ where the saddle-point of the integral
of Eq. \ref{defXq} vanishes $\kappa(q \geq q_{sat})=0$.
It is very natural to expect some relation
between the two multifractal spectra $f(\alpha)$ and $\Phi(\kappa)$,
and the possibility proposed in \cite{janssen99} is that before 
the freezing of Eq. \ref{freezing} occurs, 
the transmission should scale as 
the product of two independent
 local densities of states (Eq. \ref{rhomoments})
 \begin{eqnarray}
X(q)  = 2 \Delta(q)    \ \ \ {\rm for } \ \ q \leq q_{sat} 
\label{Xqsat}
\end{eqnarray}
We refer to \cite{janssen99} for physical 
arguments in favor of this relation. 
Equations \ref{freezing} and \ref{Xqsat} for the moments exponents
are equivalent to following relation between the two
multifractal spectra
\begin{eqnarray}
\Phi(\kappa \geq 0) = 2 \left[ f( \alpha= d+ \frac{\kappa}{2}   ) -d \right]
\label{resphikappa}
\end{eqnarray}

In this paper, our aim is to test numerically these predictions
for the statistics of the two-point transmission $T_L$
at the critical points of the Power-law Random Banded Matrix (PRBM) model,
where one parameter allows to interpolate continuously
between weak multifractality and strong multifractality.
We will also discuss the statistics of the two-point transmission
off criticality.

The paper is organized as follows. In section \ref{secmodel}, we 
introduce the PRBM model and the scattering geometry used
to define the two-point transmission. 
In Section \ref{seccriti}, we present our numerical results concerning
the multifractal statistics of the two-point transmission at criticality.
We then discuss the statistics of the two-point transmission
in the localized phase (Section \ref{secloc})
and in the delocalized phase (Section \ref{secdeloc}) respectively.
Our conclusions are summarized in section \ref{secconclusion}.
The appendices A and B contain more details on the numerical computations.

\section{ Model and observables  }

\label{secmodel}

Beside the usual short-range Anderson tight-binding model
in finite dimension $d$, other models displaying Anderson localization
have been studied,
in particular the Power-law Random Banded Matrix (PRBM) model,
which can be viewed as a one-dimensional model with long-ranged
random hopping decaying as a power-law $(b/r)^a$ of the distance $r$
with exponent $a$ and parameter $b$
(see below for a more precise definition of the model).
The Anderson transition at $a=1$ between localized ($a>1$)
 and extended ($a<1$) states
has been characterized in \cite{mirlin96} via a mapping onto a non-linear
sigma-model. The properties of the critical points at $a=1$ 
have been then much studied, in particular
the statistics of eigenvalues \cite{varga00,kra06,garcia06}, 
and the multifractality of eigenfunctions 
\cite{mirlin_evers,cuevas01,cuevas01bis,varga02,cuevas03,mirlin06},
including boundary multifractality \cite{mildenberger}.
The statistics of scattering phases, Wigner delay times
and resonance widths in the presence of one external wire 
have been discussed in \cite{mendez05,mendez06}.
Related studies concern dynamical aspects \cite{limadyn},
the case with no on-site energies \cite{lima},
and the case of power-law hopping terms in dimension $d>1$ \cite{potempa,
cuevas04,cuevas05}. In this paper, we consider the PRBM in 
a ring geometry (dimension $d=1$ with periodic boundary conditions)
in the presence of two external wires to measure the transmission properties.

\subsection{Power-law random banded matrices
 with periodic boundary conditions }

We consider $L$ sites $i=1,2,..L$ in a ring geometry
with periodic boundary conditions,
where the appropriate distance $r_{i,j}$
between the sites $i$ and $j$ is defined as \cite{mirlin_evers}
\begin{eqnarray}
r_{i,j}^{(L)} = \frac{L}{\pi} \sin \left( \frac{ \pi (i-j) }{L} \right)
\label{rijcyclic}
\end{eqnarray}
The ensemble of power-law random banded matrices of size 
$L \times L$ is then defined as follows : 
  the matrix elements $H_{i,j}$ are independent Gaussian
variables of zero-mean $\overline{H_{i,j}}=0$ and of variance
\begin{eqnarray}
\overline{ H_{i,j}^2 } = \frac{1}{1+ \left( \frac{r_{i,j}}{b}\right)^{2a}}
\label{defab}
\end{eqnarray}
The most important properties of this model are the following.
The value of the exponent $a$ determines the localization properties
\cite{mirlin96} : 
 for $a>1$ states are localized with integrable power-law tails,
whereas for $a<1$ states are delocalized.
At criticality $a=1$, states become multifractal 
\cite{mirlin_evers,cuevas01,cuevas01bis,varga02} and exponents
depend continuously of the parameter $b$, which plays a role analog
to the dimension $d$ in short-range Anderson transitions 
\cite{mirlin_evers} : the limit $b \gg 1$ corresponds to
weak multifractality ( analogous to the case $d=2+\epsilon$)
and can be studied via the mapping onto a non-linear sigma-model
 \cite{mirlin96},
whereas the case $b \ll 1$ corresponds to strong multifractality
( analogous to the case of high dimension $d$)
and can be studied via Levitov renormalization \cite{levitov,mirlin_evers}.
Other values of $b$ have been studied numerically  
\cite{mirlin_evers,cuevas01,cuevas01bis,varga02}.

\subsection{ Scattering geometry used to define to two-point transmission}

\begin{figure}[htbp]
 \includegraphics[height=6cm]{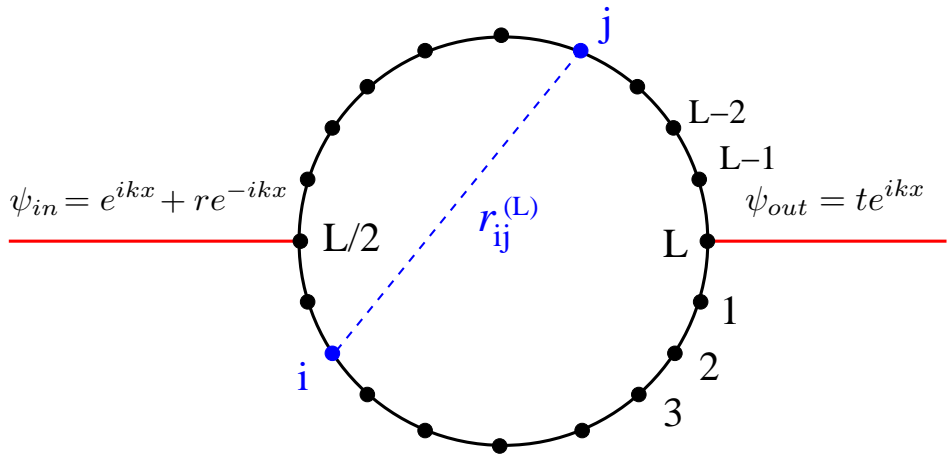}
\vspace{1cm}
\caption{ (Color on line) 
The ensemble of power-law random banded matrices of size 
$L \times L$ can be represented as a ring of $L$ sites,
where the matrix element $H_{i,j}$ between the sites $i$ and $j$
is a Gaussian variable of zero-mean $\overline{H_{i,j}}=0$
 and of variance given by Eq. \ref{defab} in terms of the distance 
 $r_{i,j}$ of Eq. \ref{rijcyclic}.
In this paper, we study the statistics of the Landauer transmission
 $T_L=\vert t \vert^2$ when an incoming wire is attached at the site $L/2$ 
and an outgoing  wire is attached at the site $L$ (see text for more details). 
  }
\label{figring}
\end{figure}

In quantum coherent problems, the most appropriate characterization 
of transport properties consists in defining a scattering problem
where the disordered sample is linked to incoming wires and outgoing wires,
and in studying the reflection and transmission coefficients.
This scattering theory definition of transport, 
first introduced by Landauer \cite{landauer},
has been often used for one-dimensional systems 
\cite{anderson_fisher,anderson_lee,luck}
and has been generalized to higher dimensionalities and multi-probe
measurements (see the review \cite{stone}).
In the present paper, 
we focus on the Landauer transmission
for the scattering problem shown on Fig. \ref{figring}:
an incoming wire is attached at the site $L/2$ 
and an outgoing  wire is attached at the site $L$. 
We are thus interested into the eigenstate $\vert \psi >$ that satisfies 
the  Schr\"odinger equation
\begin{eqnarray}
H \vert \psi > = E \vert \psi > 
\label{schodinger}
\end{eqnarray}
inside the disorder sample characterized by the random $H_{i,j}$,
and in the perfect wires
characterized by no on-site energy and by hopping unity between nearest
neighbors.
Within these perfect wires, one requires the plane-wave forms
\begin{eqnarray}
\psi_{in}(x \leq x_{L/2}) && = e^{ik (x-x_{L/2})} +r e^{- i k (x-x_{L/2})} \nonumber \\
 \psi_{out}(x \geq x_L) && = t e^{ik (x-x_{L})} 
\label{psiwires}
\end{eqnarray}
These boundary conditions define
 the reflection amplitude $r$ of the incoming wire
and the transmission amplitude $t$ of the outgoing wire.
The Landauer transmission  
\begin{eqnarray}
T \equiv  \vert t \vert^2 = 1 - \vert r \vert^2
\label{deftotaltrans}
\end{eqnarray}
is then a number in the interval $[0,1]$.
More details on the numerical computation of the transmission
in a given sample
are given in Appendix \ref{apponesample}.

To satisfy the Schr\"odinger Equation of 
Eq. \ref{schodinger} within the wires with
the forms of Eq. \ref{psiwires}, one has the following relation between
the energy $E$ and the wave vector $k$  
\begin{eqnarray}
 E=2 \cos k  
\label{relationEk}
\end{eqnarray}
To simplify the discussion, we will focus in this paper on the case of
zero-energy $E=0$ (wave-vector $k=\pi/2$)
 that corresponds to the center of the band.

In the following, we study numerically the statistical properties
of the Landauer transmission $T$ for rings of size $50 \leq L \leq 1800$
with corresponding statistics of $10.10^8 \geq n_s(L) \geq 2400$
 independent samples.
For typical values, the number $n_s(L)$ of samples is sufficient
even for the bigger sizes, whereas for the measure of multifractal
spectrum, we have used only the smaller sizes where the statistics
of samples was sufficient to measure correctly the rare events.

\section{ Statistics of the two-point transmission at criticality  ($a=1$) }

\label{seccriti}

As recalled in the Introduction,
the two-point transmission $T_L$ is 
expected to display multifractal statistics
at criticality \cite{janssen99}. We first focus
on the scaling of the typical transmission before
we turn to the multifractal spectrum and the moments of arbitrary order.

\subsection{ Typical transmission at criticality ($a=1$)}

\begin{figure}[htbp]
 \includegraphics[height=6cm]{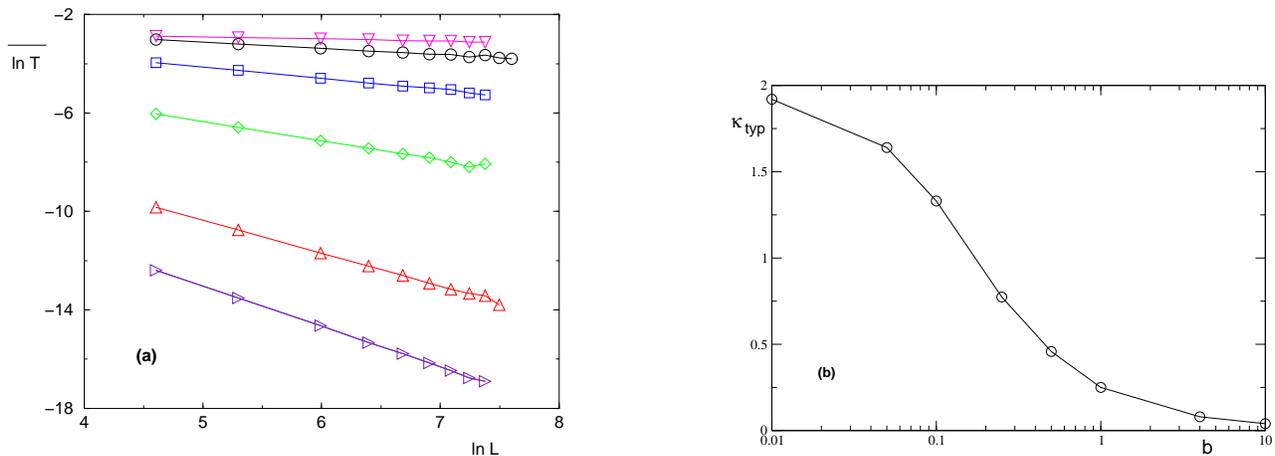}
\hspace{2cm}
 \includegraphics[height=5cm]{xmkappatypb.eps}
\vspace{1cm}
\caption{ (Color on line) 
Scaling of the typical transmission $T_L^{typ}$ at criticality $a=1$ 
for various values of the parameter $b$ :
(a) $ \ln T_L^{typ}\equiv  \overline{ \ln T_L } $
as a function of $ \ln L$ for 
various values of 
$b=4 \ (\triangledown), 1 \  (\bigcirc), 0.5 \ (\square), 0.25
\ (\Diamond), 0.1 \ (\vartriangle), 0.05 \ (\vartriangleright)$
(the two other values $b=10$ and $b=0.01$ we have studied
are not shown here for clarity) :
the slope yields the typical exponent $\kappa_{typ}(b) $
 of Eq. \ref{Ttypcriti}.
(b) critical exponent  $\kappa_{typ}(b) $ as a function of $b$. }
\label{figtranscritib1}
\end{figure}

As discussed in \cite{janssen99,evers08},
the typical transmission
\begin{eqnarray}
   T^{typ}_L \equiv e^{ \overline{ \ln T_L } } 
\label{defTtyp}
\end{eqnarray}
is expected to decay at criticality with some power-law
\begin{eqnarray}
   T^{typ}_L  
\oppropto_{L \to \infty} \frac{1}{L^{\kappa_{typ}}}
\label{defTtypcriti}
\end{eqnarray}
where the exponent $\kappa_{typ}$ 
is directly related via the relation 
\begin{eqnarray}
\kappa_{typ} = 2 (\alpha_{typ}- d)
\label{relationxtypalphatyp}
\end{eqnarray}
to the typical exponent $\alpha_{typ}$ 
that characterizes the typical weight of eigenfunctions
\begin{eqnarray}
\vert \psi ( \vec r) \vert^2_{typ} \propto \frac{1}{L^{\alpha_{typ}}}
\label{defalphatyp}
\end{eqnarray}
This typical value $\alpha_{typ}$
 corresponds to the maximum value $f(\alpha_{typ})=d$
of the multifractal spectrum $f(\alpha)$ 
introduced in Eq. \ref{nlalpha} .
(Note that in \cite{janssen99,evers08}), $\kappa_{typ}$ is denoted by $X_t$
and $\alpha_{typ}$ by $\alpha_0$. Here we have chosen to use the 
explicit notation 'typ' for clarity).

For the PRBM considered here, the dimension is $d=1$,
and critical exponents depend continuously on $b$. 
We show on Fig. \ref{figtranscritib1} (a) the $\ln T_L^{typ}$ 
as a function of $ \ln L$ : the slopes allows to measure
the exponents $\kappa_{typ}(b)$
\begin{eqnarray}
\ln T_L^{typ}(a=1,b) \equiv \overline{\ln T_L(a=1,b) }
 \oppropto_{L \to \infty} - \kappa_{typ}(b) \ln L
\label{Ttypcriti}
\end{eqnarray}
On Fig. \ref{figtranscritib1} (b), we show how
the exponent $\kappa_{typ}(b)$ depends on $b$.
The values $\kappa_{typ}(b)$ we have measured are listed in
Table \ref{tableres}, together with the
 corresponding values of $\alpha_{typ}(b)$ 
obtained via Eq. \ref{relationxtypalphatyp} : these values
of $\alpha_{typ}(b)$ 
are compatible with the values of the maxima of the multifractal
spectrum $f(\alpha)$ of critical eigenstates
measured in \cite{mirlin_evers} 
(see Fig. 2 and Fig. 6 of \cite{mirlin_evers}) and 
in \cite{mirlin06}  (see Fig. 2 and Fig. 3 of \cite{mirlin06}).

 \begin{table}[htbp]
 \centerline{
 \begin{tabular}{|l|l|l|l|l|l|l|l|l|l|l|}   \hline
 b  &  $ b \to 0$ & 0.01 & 0.05    &   0.1 & 0.25 & 0.5 & 1 & 4 & 10 & $b \to +\infty$   \\ \hline
$\kappa_{typ}(b)$   & 2 &  1.92 & 1.64  & 1.33 & 0.77 & 0.46 & 0.25 & 0.08 & 0.04 & 0  \\ \hline
 $\alpha_{typ}(b)=1+\frac{\kappa_{typ}(b)}{2} $ &  2  & 1.96 & 1.82  & 1.66 & 1.38 &  1.23 & 1.12 & 1.04 & 1.02 & 1 \\ \hline
 \end{tabular}
 }
\caption{Critical exponents as a function of $b$ :
(i) the exponent $\kappa_{typ}(b)$ characterizes
the typical transmission at criticality (see Eq. \ref{Ttypcriti}),
(ii) the corresponding value of the typical exponent
 $\alpha_{typ}=1+\kappa_{typ}(b)/2$
(see Eq. \ref{relationxtypalphatyp}) for the weight $\psi^2(\vec r)$ 
of eigenfunctions (see Eq. \ref{defalphatyp}) }
\label{tableres}
\end{table}

The two limits $b \gg 1$ and $b \ll 1$ can be understood as follows.
The case $b \gg 1$ corresponds to very weak
 multifractality with the typical exponent $\alpha_{typ} \to 1$ 
\cite{mirlin_evers,mirlin06}.  
Equation \ref{relationxtypalphatyp}
yields that the critical exponent $\kappa_{typ}$ of the typical
transmission becomes arbitrary small in the limit $b \to +\infty$
\begin{eqnarray}
\kappa_{typ}(b \to +\infty) \to 0
\label{kappaweak}
\end{eqnarray}
The opposite limit $b \ll 1$ corresponds to very strong multifractality
with the typical exponent $\alpha_{typ} \to 2$
 \cite{mirlin_evers,mirlin06}. Equation \ref{relationxtypalphatyp}
thus yields 
\begin{eqnarray}
\kappa_{typ}(b \to 0) \to 2
\label{kappastrong}
\end{eqnarray}

\subsection{ Multifractal spectrum $\Phi_b(\kappa)$ 
with termination at $\kappa=0$  }

\begin{figure}[htbp]
 \includegraphics[height=6cm]{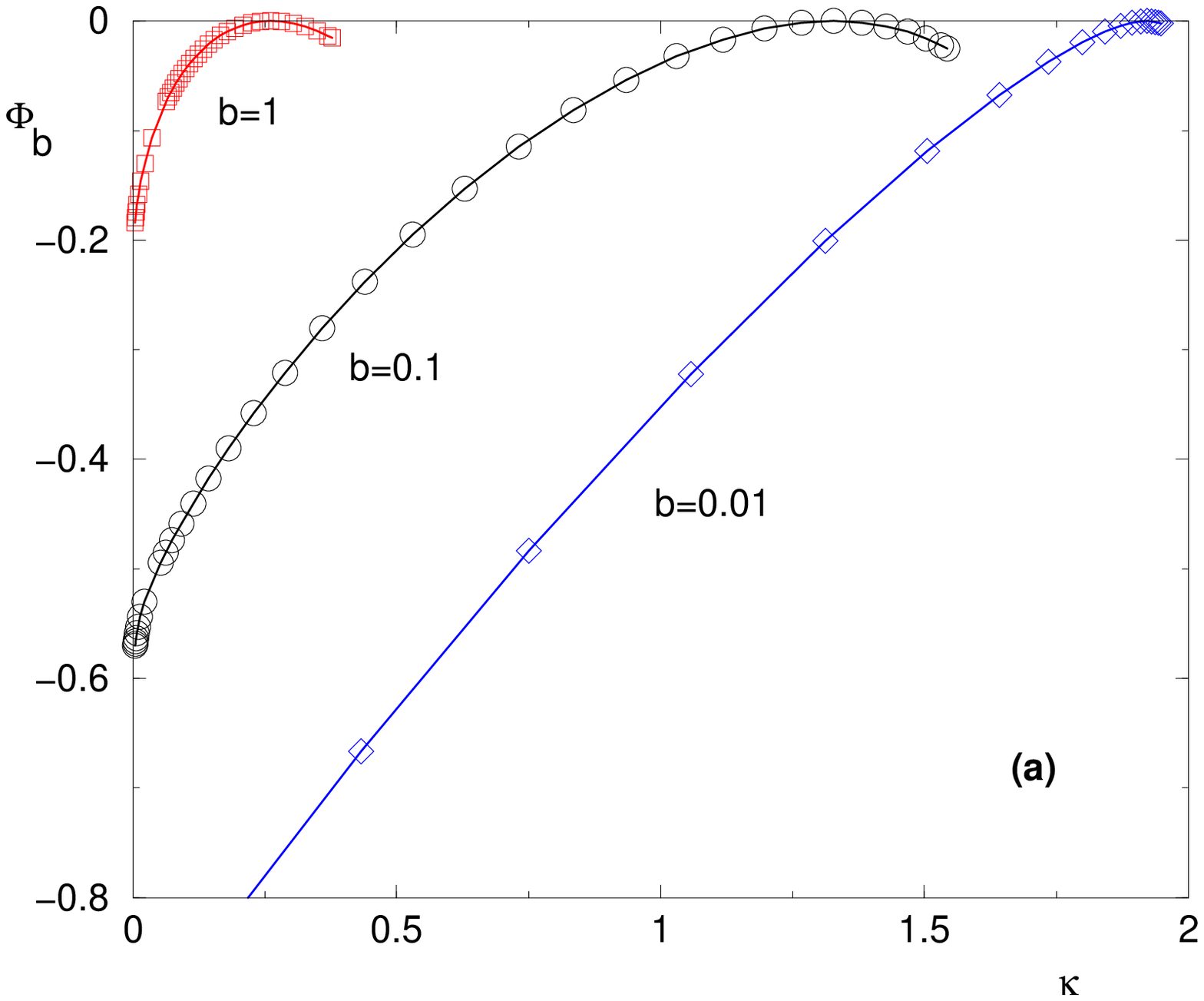}
\hspace{2cm}
 \includegraphics[height=6cm]{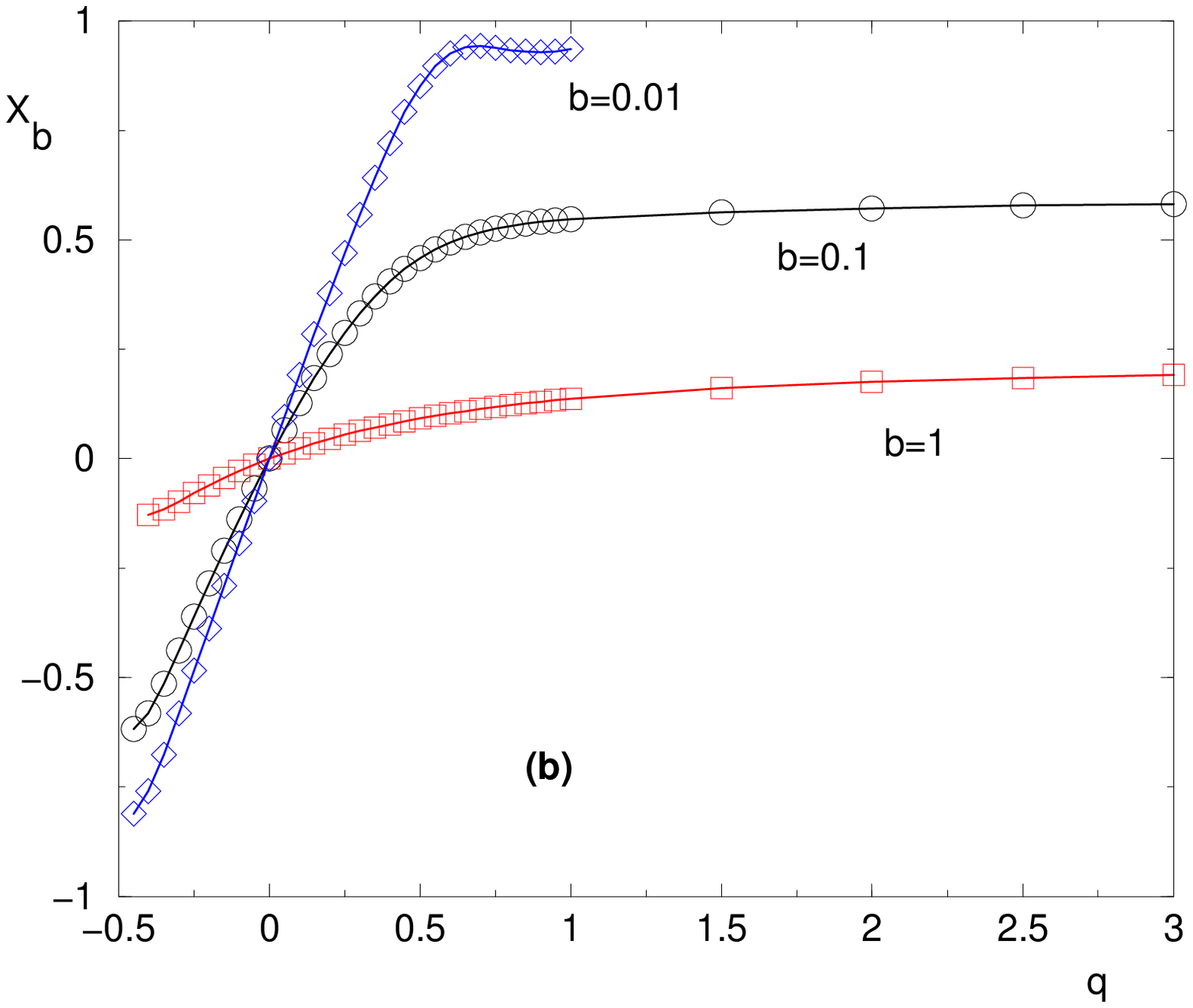}
\vspace{1cm}
\caption{ (Color on line) 
Multifractal statistics of the critical two-point transmission 
for three values $b=1$, $b=0.1$ and $b=0.01$ : 
(a) the multifractal spectra $\Phi_b(\kappa)$
reach their maximum $\Phi_b(\kappa_{typ}(b))=0$ at the typical values $\kappa_{typ}(b)$ and exactly terminate at  finite values $\Phi_b(\kappa=0)$.
(b) the corresponding moments exponents $X_b(q)$ become completely
frozen for $q \geq q_{sat}$ : 
$X_b(q \geq q_{sat}) =  - \Phi_b(\kappa=0)$.  }
\label{figmultif3valuesb}
\end{figure}

As recalled in the introduction,
the statistics of the two-point transmission 
is expected to be multifractal at criticality, as a consequence of the 
multifractal character of critical eigenfunctions \cite{janssen99}.
For the PRBM, the multifractal 
spectrum $\Phi_b(\kappa)$ of Eq. \ref{phikappa}
will depend continuously on the parameter $b$
\begin{equation}
{\rm Prob}\left( T_L \sim L^{-\kappa}  \right) dT
\oppropto_{L \to \infty} L^{\Phi_b(\kappa) } d\kappa
\label{phikappab}
\end{equation}
Since it describes a probability, the multifractal spectrum 
satisfies $\Phi_b(\kappa) \leq 0$, and 
the maximal value $\Phi_b(\kappa) = 0$
is reached only for the typical value $\kappa=\kappa_{typ}(b)$
 discussed above
\begin{equation}
\Phi_b(\kappa_{typ}(b)) =0
\label{phikappabtyp}
\end{equation}
The relation of Eq. \ref{relationxtypalphatyp}
 between the two typical exponents
$\kappa_{typ}(b)$ and $\alpha_{typ}(b)$ is expected to come from the
more general relation of Eq. \ref{resphikappa}
between the two multifractal spectra $\Phi_b(\kappa)$ and $f_b(\alpha)$
\begin{eqnarray}
\Phi_b(\kappa \geq 0) = 2 \left[ f_b( \alpha= 1+ \frac{\kappa}{2}   ) - 1 \right]
\label{resphikappab}
\end{eqnarray}
An essential property of the spectrum $\Phi_b(\kappa)$ is that it 
exists only for $\kappa \geq 0$ as a consequence of the physical bound 
$T_L \leq 1$, so that it terminates at $\kappa=0$ at the finite value
\begin{eqnarray}
\Phi_b(\kappa = 0) = 2 \left[ f_b( \alpha= 1 ) - 1 \right]
\label{termination}
\end{eqnarray}

To measure numerically the multifractal spectrum $\Phi_b(\kappa)$,
we have used the standard method based on $q$-measures of Ref. \cite{Chh}
(see more details in Appendix \ref{appmultif}).
To show how the parameter $b$ allows to interpolate
 between weak multifractality and strong multifractality, we compare on  
Fig. \ref{figmultif3valuesb} (a)
the multifractal spectra $\Phi_b(\kappa)$ for the three values 
$b=1$, $b=0.1$ and $b=0.01$.
For instance for the value $b=0.1$,
the termination value we measure $\Phi_b(\kappa = 0) \sim - 0.58$
is in agreement via Eq. \ref{termination}
with the value $f_b( \alpha= 1 ) \sim 0.71$ 
of Fig. 6 of \cite{mirlin_evers} and Fig. 2 of \cite{mirlin06}.

\subsection{ Freezing transition of the moments exponents $X_b(q)$  }

As usual, the multifractal statistics of Eq. \ref{phikappab}
has for consequence that the moments of arbitrary order $q$
\begin{equation}
\overline{T_L^q} \sim \int d\kappa L^{\Phi_b(\kappa) -q \kappa }
\oppropto_{L \to \infty} L^{-X_b(q)}
\label{defXqb}
\end{equation}
are governed by non-trivial exponents $X_b(q)$
that can be obtained via the saddle-point calculation
\begin{eqnarray}
 - X_b(q)  = \opmax_{ \kappa \geq 0} \left[ \Phi_b(\kappa)- \kappa q \right]
\label{saddleb}
\end{eqnarray}
As long as the saddle-point value satisfies $\kappa(q) \geq 0$,
$X_b(q)$ can be obtained via the usual 
Legendre transform formula
\begin{eqnarray}
 q && =\Phi_b'(\kappa) \nonumber \\
 X_b(q) && = \kappa q - \Phi_b(\kappa)
\label{legendreb}
\end{eqnarray}
However above some threshold $q_{sat}$, the saddle-point value
will saturate to the boundary value 
\begin{eqnarray}
\kappa(q \geq q_{sat}) =0
\label{colzero}
\end{eqnarray}
and the exponent $X(q)$ will saturate to the value
\begin{equation}
X_b(q \geq q_{sat}) = X_b( q_{sat}) = 
- \Phi_b(\kappa=0) = 2 (1- f_b(\alpha=1))
\label{xqfreezing}
\end{equation}
This freezing phenomenon of $X(q)$ at $q_{sat}$
 predicted in \cite{janssen99} means 
that all moments of order $q \geq q_{sat}$ are dominated
by the rare events corresponding to a finite transmission
 $T \simeq 1$, whose measure behaves as  $L^{\Phi(\kappa=0)}$.

We show on Fig. \ref{figmultif3valuesb} (a)
the moments exponents $X_b(q)$ for the three values 
$b=1$, $b=0.1$ and $b=0.01$.
For instance for $b=0.1$,
the freezing value $X_b(q \geq q_{sat}) \sim 0.58 $ 
corresponds to the termination value 
$\Phi_b(\kappa=0) \sim -0.58$ of Fig. \ref{figmultif3valuesb} (a).

It turns out that for Anderson transitions, a special symmetry of the
multifractal spectrum $f(\alpha)$
has been proposed (see \cite{mirlin06,vasquez} and references therein) 
that relates the regions $\alpha \leq d$ and $\alpha \geq d$
via the relation $f(2d-\alpha)=f(\alpha)+d-\alpha$.
This symmetry then 
  fixes the value of $q_{sat}$ where $\kappa(q_{sat})=0$ or equivalently
$\alpha(q_{sat})=d$ to be exactly
\begin{eqnarray}
q_{sat}= \frac{1}{2}
\label{qsatdemi}
\end{eqnarray}
Numerically, it is difficult to
measure precisely the value $q_{sat}$ where the exponents $X_b(q)$
become completely frozen as a consequence of finite-size corrections
around this phase transition point for the $X(q)$,
as already found for the quantum Hall transition
in \cite{janssen99}. However Fig. \ref{figmultif3valuesb} (b) shows
that in the limit of strong multifractality ($b=0.01$), the numerical
saturation value is not far from the theoretical prediction of
 Eq. \ref{qsatdemi}.

\section{ Statistics of the two-point transmission in the 
localized phase ($a>1$)}

\label{secloc}

\subsection{ Typical transmission in the 
localized phase}

\begin{figure}[htbp]
 \includegraphics[height=6cm]{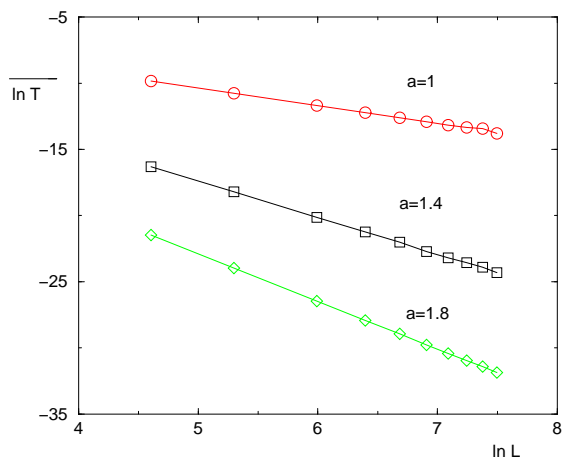}
\caption{(Color on line) 
Typical two-point transmission in the localized phase $a>1$.
The power-law decay of Eq. \ref{defTtyploc} is checked here
for the case $b=0.1$ and the two values $a=1.4$, $a=1.8$ :
the slopes of this log-log plot are of order $2 a \sim 2.8$ and
  $2 a \sim 3.6$.
For comparison, we also show the critical data for $a=1$ of slope
$\kappa_{typ}(b=0.1) \sim 1.33$.
}
\label{figtyploc}
\end{figure}

In usual short-range models, the localized phase
is characterized by exponentially localized wavefunctions,
whereas in the presence of power-law hoppings, 
localized wavefunction can only decay with power-law integrable tails.
For the PRBM, it is moreover expected that the asymptotic
decay is actually given exactly by the power-law of 
Eq. \ref{defab} for the hopping term defining the model \cite{mirlin96} :
$\vert \psi (r) \vert^2_{typ} \sim 1/r^{2a}$.
As a consequence in the localized phase $a>1$, one expects 
the typical decay  
\begin{eqnarray}
   T^{typ}_L(a>1) \oppropto_{L \to \infty} \frac{1}{L^{2a}}
\label{defTtyploc}
\end{eqnarray}
As shown on Fig. \ref{figtyploc}, we have checked
 this power-law decay of the typical transmission
for the case $b=0.1$ and the two values $a=1.4$, $a=1.8$.

\subsection{ Histogram of $(\ln T_L)$ in the localized phase ($a>1$)  }
 
\begin{figure}[htbp]
 \includegraphics[height=6cm]{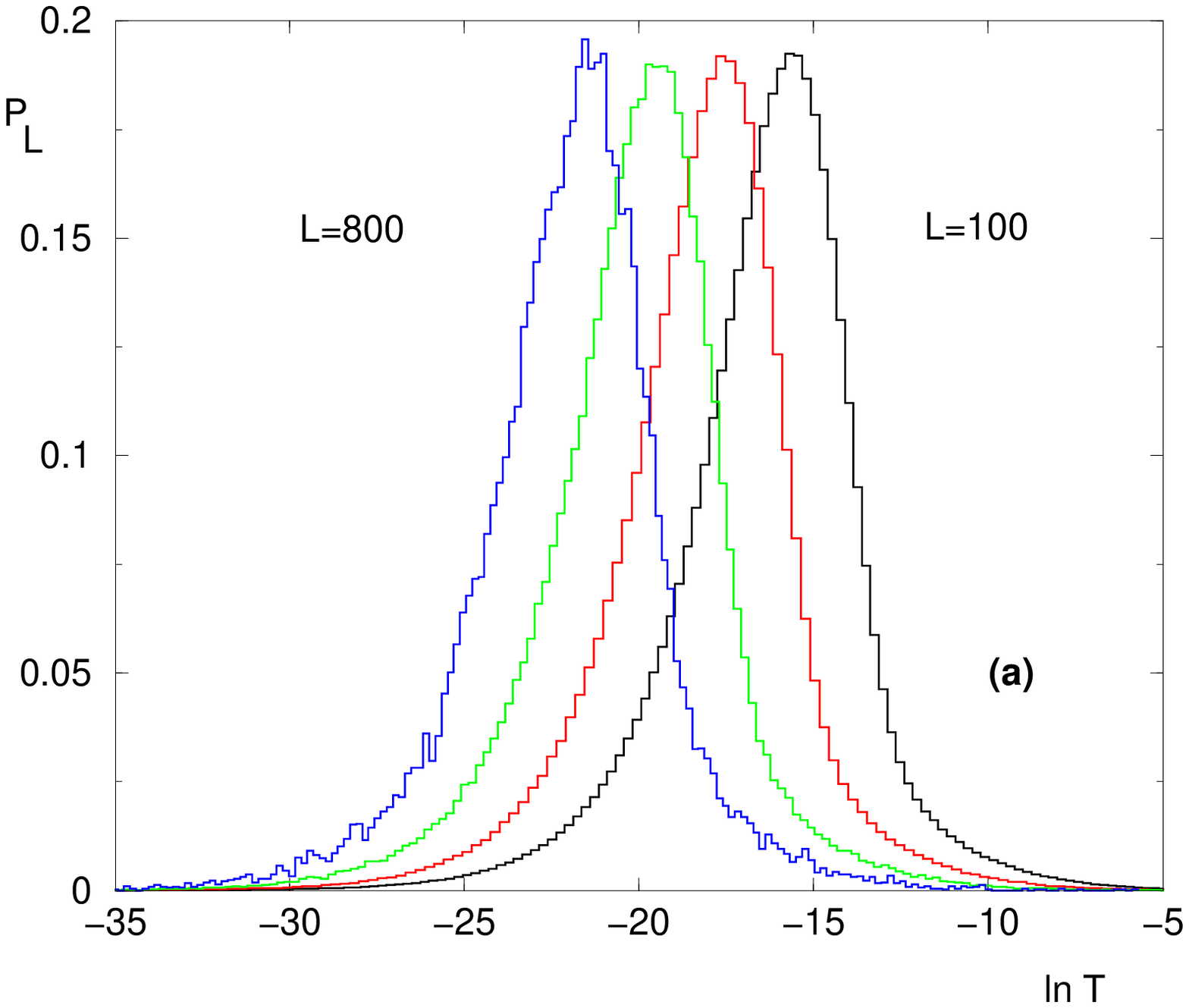}
\hspace{2cm}
 \includegraphics[height=6cm]{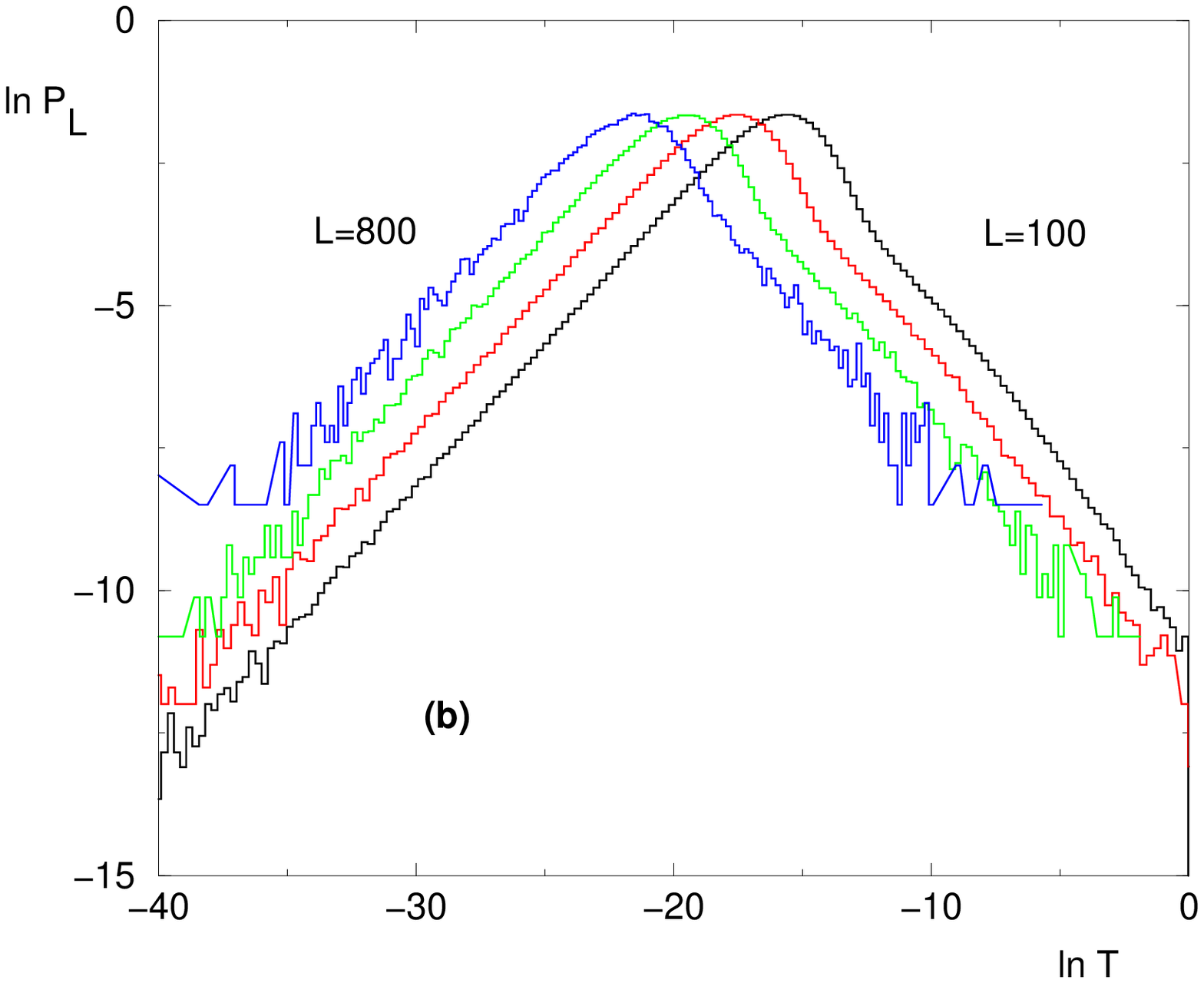}
\caption{(Color on line) Histogram of of the two-point transmission
 in the localized phase $a>1$ : example for $b=0.1$ and $a=1.4$ :
 (a) histograms $ P_L$ of $\ln T_L$ for the four sizes $L=100,200,400,800$
 (the histograms for bigger sizes
 are more noisy and are not shown for clarity) 
(b) same data in log scale to exhibit the tails discussed in the text.}
\label{fighistoloc}
\end{figure}

We show on Fig. \ref{fighistoloc} the histograms $ P_L$ 
of $\ln T_L$ for the four sizes $L=100,200,400,800$
 in the localized phase $a=1.4$
for the case $b=0.1$ : these data seem to indicate
that as $L$ grows, the probability distribution shifts to the left
while keeping a fixed shape. 
This means that the relative variable $u=(\ln T_L- \ln T_L^{typ})$
with respect to the typical value discussed above 
(see Eq. \ref{defTtyploc})
remains a finite variable as $L \to +\infty$.
In addition, the left tail is governed by the exponent $\alpha \sim 0.5$ 
\begin{eqnarray}
\ln P_{\infty}^{loc}(u =(\ln T_L- \ln T_L^{typ}) ) 
\opsimeq_{u \to - \infty} \frac{1}{2} u 
\label{slopeloc}
\end{eqnarray}
or equivalently after the change of variable ${\cal T} = \frac{T_L}{T_L^{typ}}=e^u$
\begin{eqnarray}
 P_{\infty}^{loc}({\cal T} = \frac{T_L}{T_L^{typ}})  \oppropto_{{\cal T} \to 0}
 \frac{1}{{\cal T}^{1/2}}
\label{locbis}
\end{eqnarray}

These properties can be understood by the following simple argument.
In the localized phase $a>1$, one may assume that for large $L$,
the transmission $T_L$ is dominated by the direct hopping term 
$ H_{L/2,L} $ (see Fig. 1)
\begin{eqnarray}
T_L (a>1) \oppropto_{L \to \infty} \vert H_{L/2,L} \vert^2 
\oppropto_{L \to \infty} \frac{x^2}{L^{2a}} 
\label{directhopping}
\end{eqnarray}
where $x$ is a Gaussian variable of zero mean and variance unity 
(see Eq. \ref{defab}). 
In particular, its probability density is finite near 
the origin $P(x =0) >0$. The change of variable 
\begin{eqnarray}
{\cal T} = \frac{T_L}{T_L^{typ}} \propto x^2
\label{changevar}
\end{eqnarray}
then yields the power-law of Eq. \ref{locbis}.

\section{ Statistics of the two-point 
transmission in the delocalized phase ($a<1$)}

\label{secdeloc}

\subsection{ Typical transmission in the delocalized phase  }

In the delocalized phase, the eigenfunctions 
are not multifractal anymore, but monofractal with the 
single value $\alpha_{deloc}=d$ for the weight
As a consequence, the typical transmission
is expected to remain finite as $L \to +\infty$ \cite{janssen99}
(in Eq. \ref{relationxtypalphatyp}, the case $\alpha_{typ}=d$ yields 
$\kappa_{typ} = 0$)
\begin{eqnarray}
   T^{typ}_L 
\oppropto_{L \to \infty} T^{typ}_{\infty} >0
\label{defTtypdeloc}
\end{eqnarray}
The two-point transmission is thus a good order parameter
of the transport properties \cite{janssen99}.

\subsection{ Histogram of $(\ln T_L)$ in the delocalized phase $a<1$  }

\begin{figure}[htbp]
 \includegraphics[height=6cm]{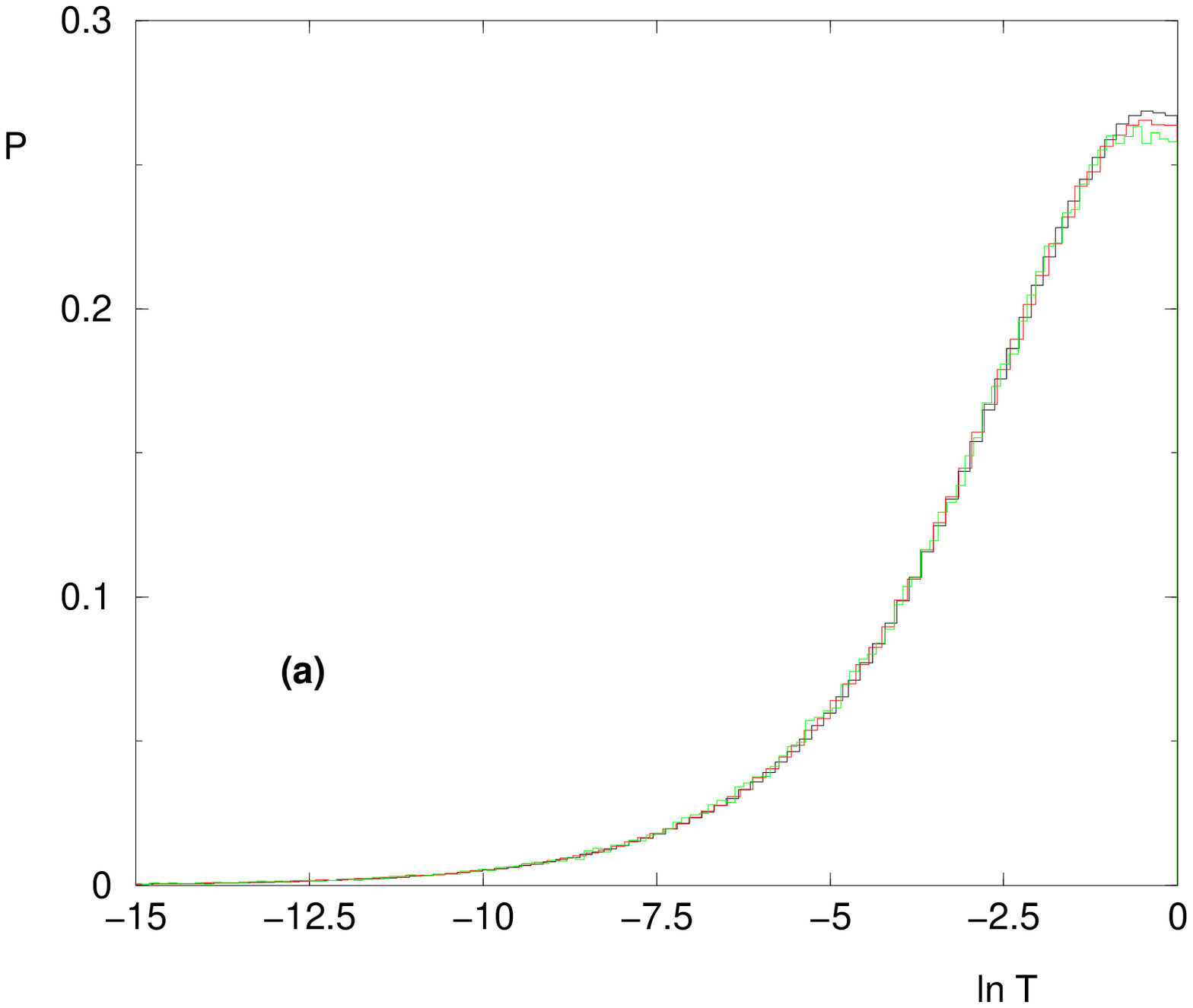}
\hspace{2cm}
 \includegraphics[height=6cm]{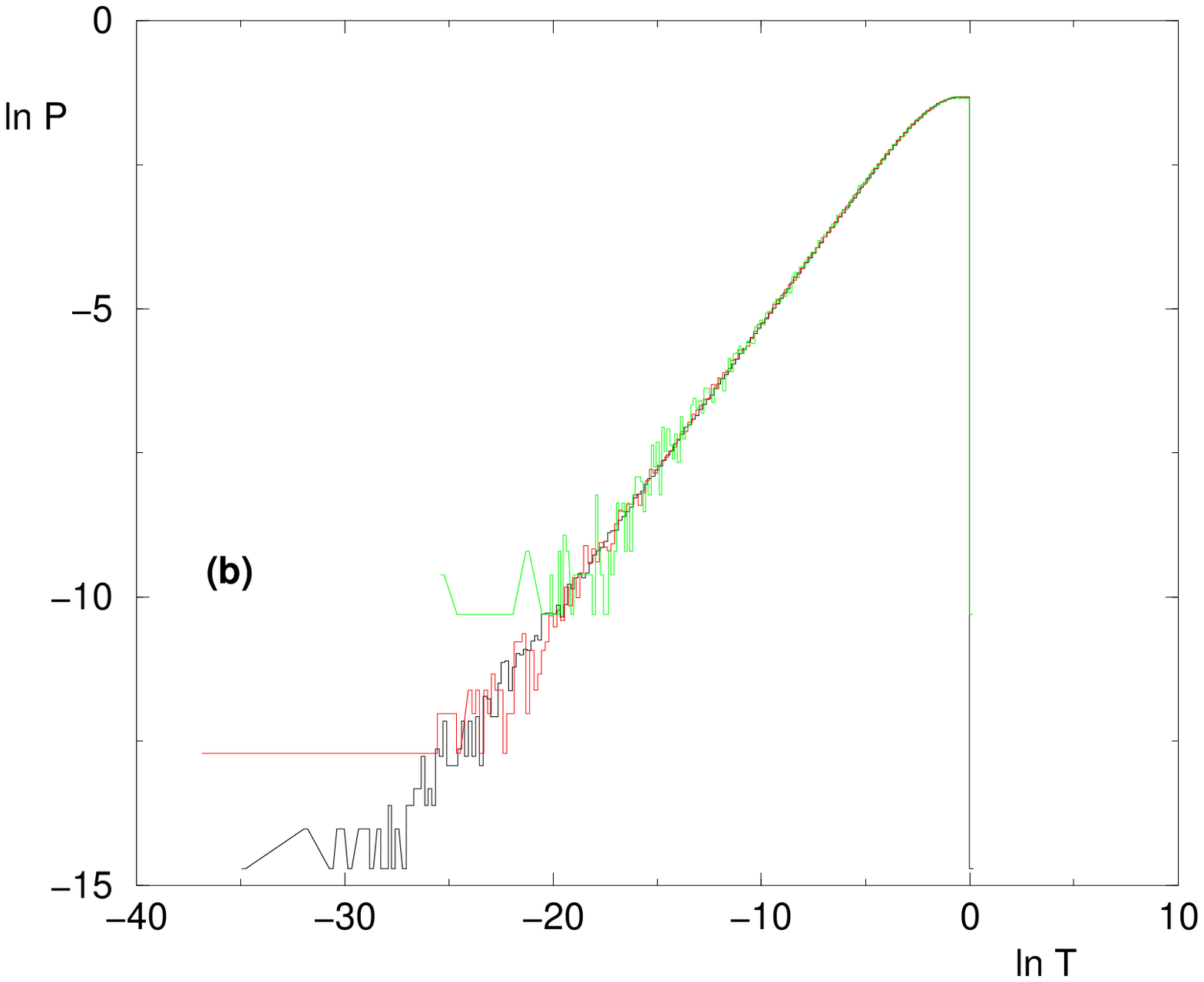}
\caption{(Color on line) Histogram of of the two-point transmission
 in the delocalized phase $a<1$ : example for $b=1$ and $a=0.75$
 (a) the histograms of $\ln T_L$ for the three sizes $L=100,200,400$
coincide (the histograms for bigger sizes are more noisy and are not shown for clarity)
(b) same data in log scale to exhibit the tail (see Eq. \ref{slopedeloc})}
\label{fighistodeloc}
\end{figure}

We find that the whole probability distribution
$P_L(\ln T)$ converges for large $L$ towards a fixed distribution
$P_{\infty}^{deloc}(\ln T)$. As an example, we show on Fig. \ref{fighistodeloc}
the histograms of $(\ln T_L)$ for three sizes $L=100,200,400$ 
concerning the case $b=1$ and $a=0.75$.
These histograms stops at $\ln T=0$ 
as a consequence of the bound $T \leq 1$. 
In the region of very small transmission $\ln T \to -\infty$
the log-log plot of Fig. \ref{fighistodeloc} (b) indicates
the same form as in Eq. \ref{slopeloc}
\begin{eqnarray}
\ln P_{\infty}^{deloc}(\ln T) 
\oppropto_{\ln T \to - \infty} \frac{1}{2} \ln T 
\label{slopedeloc}
\end{eqnarray}
or equivalently after a change of variable
\begin{eqnarray}
 P_{\infty}^{deloc}( T)  \oppropto_{T \to 0} \frac{1}{T^{1/2}}
\label{delocbis}
\end{eqnarray}
As explained around Eq. \ref{changevar}, this power-law behavior
simply means that the transmission can be written as the square
$T \simeq x^2$ of some random variable $x$ that has a finite
probability density at the origin $P(x=0)>0$.
This means that all negative moments of order $q \leq -1/2$ 
actually diverge in the delocalized phase
\begin{eqnarray}
\int_0^1 dT T^q P_{\infty}^{deloc}( T)  \opsimeq_{q \leq -1/2} +\infty
\label{delocmonneg}
\end{eqnarray}

\section{ Conclusions and perspectives }

\label{secconclusion}

In this paper, we have studied numerically the statistics of
the two-point transmission $T_L$ as a function of the size $L$
in the PRBM model that depends on two parameters $(a,b)$ :

(i) in the delocalized phase ($a<1$), we have found that
the probability distribution
of $T_L$ converges for $L \to +\infty$
towards a law $P_{\infty}^{deloc}( T)$ presenting the power-law
of Eq. \ref{delocbis} for $T \to 0$.

(ii) in the localized phase ($a>1$), 
we have found that the probability distribution
$ P_L^{loc} (\ln T_L)$ keeps a fixed shape 
around the typical value $\ln T^{typ} =\overline{ \ln T_L}$ as $L$ grows,
and the typical value $T^{typ}_L$ decays only as the power-law of
Eq. \ref{defTtyploc} as a consequence of the presence of power-law hoppings.

(iii) exactly at criticality ($a=1$), the statistics of the two-point
transmission $T_L$ is multifractal :  we have measured
the multifractal spectra $\Phi_b(\kappa)$ as well as the moments
exponents $X_b(q)$ for various values of the parameter $b$
that allows to interpolate between weak multifractality
and strong multifractality.
We have tested in detail the various expectations of Ref. \cite{janssen99}
concerning the termination of $\Phi_b(\kappa)$ at $\kappa=0$,
the freezing of $X_b(q)$ above some value $q \geq q_{sat}$,
and the relations with the singularity spectrum $f_b(\alpha)$ 
of individual critical eigenstates.

To finish, we should stress that the relation of Eq. \ref{resphikappa}
relates the transmission between two 'bulk points'
to the 'bulk multifractal spectrum' $f(\alpha)$.
In localization models, it is however more usual to attach leads
to the boundaries of the disordered sample : then the statistics 
of the two-point transmission is related to the 'surface multifractal
spectrum' as will be discussed in more details elsewhere \cite{us_many}.
We will also discuss in \cite{us_many} the statistical
properties of the transmission 
for various scattering geometries involving a large number of wires.

\appendix

\section{ Computation of the two-point transmission in each sample}

\label{apponesample}

To compute the transmission of a given sample via Eq. \ref{deftotaltrans},
we have to solve the Sch\"odinger problem of Eq. \ref{schodinger}
with the scattering boundary conditions of Eq. \ref{psiwires}.
This can be decomposed in two steps as follows.

\subsection{ Recursive elimination of the 'interior sites' }

The first step consists in the iterative elimination of the 'interior sites',
i.e. of all the sites not connected to the external wires 
(see Fig. \ref{figring}). To eliminate the site $i_0$, one uses 
the Sch\"odinger Eq. \ref{schodinger} projected on this site
\begin{eqnarray}
E \psi(i_0)= H_{i_0,i_0} \psi(i_0) +\sum_j H_{i_0,j} \psi(j)
\label{psii0}
\end{eqnarray}
to make the substitution
\begin{eqnarray}
 \psi(i_0)=  \frac{1}{E-H_{i_0,i_0}} \sum_j H_{i_0,j} \psi(j)
\label{psii0elim}
\end{eqnarray}
in all other remaining equations.
Then from the point of view of remaining sites, 
the hoppings are renormalized according to
\begin{eqnarray}
H_{i,j}^{new} = H_{i,j} + \frac{H_{i,i_0} H_{i_0,j} }{E-H_{i_0,i_0}}
\label{rulev}
\end{eqnarray}
This procedure is stopped when the only remaining sites
are the two sites $L/2$ and $L$ connected to the external wires 
(see Fig. \ref{figring}) : the three real remaining parameters
are the renormalized hopping ${\tilde H}_{L/2,L}$ and
the two renormalized on-site energies ${\tilde H}_{L/2,L/2}$,
 ${\tilde H}_{L,L}$

\subsection{ Effective scattering problems for the two boundary sites }

The second step consists in solving the scattering problem 
for the two boundary sites $L/2$ and $L$ connected to the external wires 
(see Fig. \ref{figring}) with
the renormalized parameters obtained above.
The Sch\"odinger Eq. \ref{schodinger} projected on the boundary
sites $L/2$ and $L$ read
\begin{eqnarray}
E \psi(x_{L/2}) && = {\tilde H}_{L/2,L/2} \psi(x_{L/2}) + \psi(x_{L/2}-1)
+ {\tilde H}_{L/2,L}\psi(x_{L}) \nonumber \\
E \psi(x_{L}) && = {\tilde H}_{L,L} \psi(x_{L}) + \psi(x_{L}+1)
+ {\tilde H}_{L,L/2}\psi(x_{L/2})
\label{psibounddaries}
\end{eqnarray}
The boundary conditions of Eq. \ref{psiwires}
fixes the following ratio on the outgoing wire
\begin{eqnarray}
\frac{ \psi(x_{L}+1)}{\psi(x_{L}) } = e^{ik}
\label{outratio}
\end{eqnarray}
The following ratio 
\begin{eqnarray}
R \equiv \frac{\psi(x_{L/2}-1)}{ \psi(x_{L/2})}
\label{riccdef}
\end{eqnarray}
concerning the incoming wire can be then computed
in terms of the three real renormalized parameters
\begin{eqnarray}
R = E - {\tilde H}_{L/2,L/2}
 - \frac{{\tilde H}_{L,L/2}^2}{E-\left({\tilde H}_{L,L}+ e^{ik} \right)}
\label{ricc}
\end{eqnarray}
The reflexion coefficient $r$ of Eq. \ref{psiwires}
is then obtained as
\begin{eqnarray}
r = \frac{R-e^{-ik} }{e^{ik}-R}
\label{reflex}
\end{eqnarray}
yielding the transmission of Eq. \ref{deftotaltrans}.

\section{ Computation of the multifractal spectrum over the samples}

\label{appmultif}

To measure numerically the multifractal spectrum $\Phi(\kappa)$
of Eq. \ref{phikappa} that characterizes the statistics of the transmission
$T_L$ over the samples of size $L$, 
we have used the standard method based on $q$-measures of Ref. \cite{Chh}.
More precisely, for various sizes $L$, we have measured 
the transmission $T_L(i)$ for a number $n_s(L)$ of independent samples $(i)$.
Then for various values of $q$, we have computed the moments
of Eq. \ref{defXq}
\begin{equation}
\overline{T_L^q} = \frac{1}{n_s(L)} \sum_{i=1}^{n_s(L)} \left[ T_L(i) \right]^q
\label{numemomq}
\end{equation}
to extract the moments exponents $X(q)$ as the slope of
the log-log plot
\begin{equation}
\ln \left( \overline{T_L^q} \right)
\oppropto_{L \to \infty} -X(q) \ln L
\label{mesXq}
\end{equation}
We have also computed the auxiliary observables
\begin{equation}
K_L(q) = \frac{ \displaystyle \sum_{i=1}^{n_s(L)} \left[ T_L(i) \right]^q 
\left( - \ln T_L(i) \right)  }
{\displaystyle \sum_{i=1}^{n_s(L)} \left[ T_L(i) \right]^q }
\label{auxiK}
\end{equation}
and
\begin{equation}
F_L(q) = q K_L(q) + \ln \left( \overline{T_L^q} \right) 
\label{auxiF}
\end{equation}
to obtain $\kappa(q)$ and $\Phi(\kappa(q))$ as the slopes
of 
\begin{equation}
 K_L(q) \oppropto_{L \to \infty} \kappa(q) \ln L
\label{slopeK}
\end{equation}
and
\begin{equation}
 F_L(q) \oppropto_{L \to \infty} \Phi(\kappa(q)) \ln L
\label{slopeF}
\end{equation}
This yields a parametric plot in $q$ of the
multifractal spectrum $\Phi(\kappa)$ :
on Figure \ref{figmultif3valuesb} (a), each circle 
of coordinate $(\kappa,\Phi(\kappa))$ corresponds to a value of $q$.

\end{document}